\documentclass[twocolumn,showpacs,prl]{revtex4}
\usepackage{amssymb}
\usepackage{graphicx}
\usepackage{dcolumn}
\usepackage{bm}

\begin{document}

\title{Control of laser wake field acceleration by plasma density profile}
\author{A.~Pukhov }
\email{pukhov@tp1.uni-duesseldorf.de}
\affiliation{Institut fur Theoretische Physik I,
  Heinrich-Heine-Universitat Duesseldorf, 40225 Duesseldorf, Germany}
\author{I.~Kostyukov}
\affiliation{Institute of Applied Physics, Russian Academy of Science, 46 Uljanov St. 
603950 Nizhny Novgorod, Russia}
\date{\today}
\begin{abstract}
We show that both the maximum energy gain and the accelerated beam
quality can be efficiently controlled by the plasma density
profile. Choosing a proper density gradient one can uplift the
dephasing limitation. When a periodic wake field is exploited, the
phase synchronism between the bunch of relativistic particles and the
plasma wave can be maintained over extended distances due to the plasma
density gradient. Putting electrons into the $n-$th wake period behind
the driving laser pulse, the maximum energy gain is increased by the
factor $2\pi n$ over that in the case of uniform plasma. The
acceleration is limited then by laser depletion rather than by
dephasing. Further, we show that the natural energy spread of
the particle bunch acquired at the acceleration stage can be
effectively removed by a matched deceleration stage, where a larger
plasma density is used.
\end{abstract}

\maketitle

Plasma-based schemes of electron acceleration have recently
demonstrated impressive progress. Quasimonoenergetic electron bunches
with the energy up to $1$~GeV and with the charge of $50$~pC have been
generated in experiments \cite{Leemans2006}. All-optical methods for
control of the bunch parameters have been developed also \cite{Faure2006}. It
is generally believed that electrons in these experiments have been
accelerated in the ``Bubble regime'' \cite{Pukhov2002}. At the same
time, a significant advance in plasma profile engineering  has been
achieved to make laser plasma interaction more efficient: plasma
capillaries for laser guiding \cite{hooker}, fabrication of corrugated plasma
structures \cite{layer}, plasma machining \cite{hsieh}.

One of the main limitations on energy gain in laser-plasma
accelerators comes from the dephasing. The velocity of relativistic
electrons is slightly higher than the phase velocity of the wake,
which is determined by the group velocity of the 
driving laser pulse. The accelerated electrons slowly outrun the
plasma wave and leave the accelerating phase. 

The limitation caused by the dephasing can be overcome by employing a
proper plasma gradient \cite{Katsouleas1986}. The profile of the
plasma density should be such that the advance of the accelerated
electrons matches the advance of the plasma wave. The equation for
plasma density profile in the 1D configuration is 

\begin{equation}
\frac{d}{dx}\left( \frac{\Phi _{n}}{\omega _{p}(x)}\right) \simeq 1-\frac{c}{v_{gr}},  
\label{equation1}
\end{equation}

\noindent where $\omega _{p}^{2}(x)=4\pi e^{2}n(x)/m$ is the squared
plasma frequency, $n(x)$ is the plasma density, 
$\Phi _{n}= \mbox{const} $ is the phase of ultrarelativistic electrons
trapped $n$ plasma wavelengths ($\lambda _{p}=2\pi c/\omega _{p}$)
behind the laser pulse, $v_{gr}$ is the group velocity of the laser
pulse, $c$ is the speed of light, $e$ and $m$ are the electron charge
and mass respectively. For weakly relativistic laser pulses with
$a=eA/mc^{2}\ll 1$ and for rarefied plasmas $n/n_{c}\ll 1$ we can
assume $v_{gr}/c\simeq 1-n(x)/ 2n_{c} $ and $\gamma \gg
\gamma _{gr}$, where $\gamma $ is the relativistic gamma-factor of the
accelerated electrons, $\gamma _{gr}^{2}=1-v_{gr}^{2}/c^{2}$, $\
n_{c}=m\omega ^{2}/4\pi e^{2}$ is the critical plasma density and
$\omega $ is the laser frequency. The solution of
Eq.~(\ref{equation1}) for the phase synchronism in laser wake field
acceleration is  

\begin{eqnarray}
n(x) &=&\frac{n_{0}}{\left( 1-x/L_{inh}\right) ^{2/3}},  
\label{solution1} \\
L_{inh} &=&\frac{c}{\omega }\left( \frac{n_{0}}{n_{c}}\right) ^{-3/2}\frac{2\Phi _{n}}{3},  
\label{linh}
\end{eqnarray}

\noindent
where $n_{0}=n(x=0)$. It follows from Eq.~(\ref{solution1}) that the
plasma density increases along the pulse propagation and the
acceleration distance is limited by $L_{inh}$ since the plasma density
goes to infinity at $x=L_{inh}$. 

   \begin{figure}
   \includegraphics[width=5.5cm]{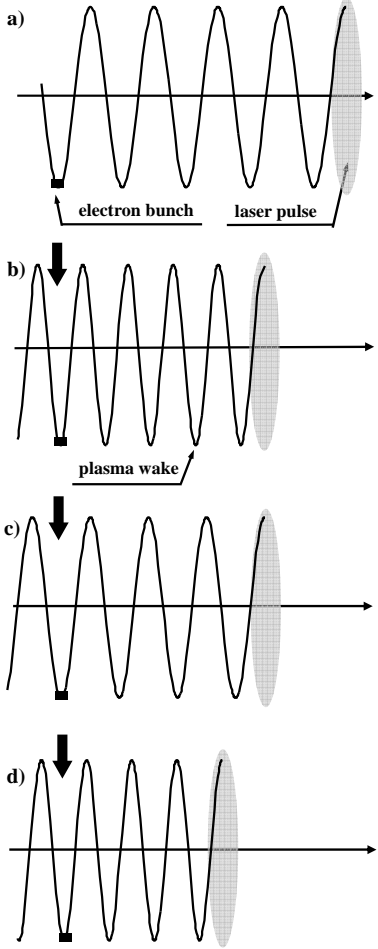}
   \caption
    {
        \label{fig1}
Electron acceleration in plasma layers in the frame of light speed
   (schematically). The ultra-relativistic electron bunch (black
   rectangle) is always located at the peak value of the accelerating field. 
    }
   \end{figure}

Consider a short circularly polarized laser pulse with the Gaussian
envelope $a^2 (\xi )=a_0^2 \exp \left( -\tau ^{2}/T^{2}\right) $,
where $\tau =t-\int^{x}dx^{\prime }/v_{gr}(x^{\prime })$. For
simplicity, the ion dynamics, the thermal motion and the transverse
dynamics of plasma electrons are neglected. It is also assumed that
the accelerated electrons do not affect 
plasma wake structure. The accelerating force on the relativistic
electrons in the plasma wake can be presented in the form  
\cite{Esarey1996} $ F_{x}=- \left( \sqrt{\pi } /2 \right) a_0^{2}mc\omega
_{p}^{2}(x)T\exp \left[ - \omega _{p}^{2}(x)T^{2} / 4 \right] \cos
\Phi (x) $ ,  where $\Phi (x) =\omega _{p}(x)\tau (x)\simeq
\sqrt{n(\xi )/n_{c}}\int n(\xi )/\left( 2n_{c}\right) d\xi $ and $\xi
=\omega x/c$. For $\cos \Phi =-1$ the accelerating force achieves a
maximum value of $F_{x}\simeq mc\omega _{p} \sqrt{\pi /2}a_{0}^{2}\exp
(-1/2)$ when $T=\sqrt{2}/\omega _{p}$. Therefore for the given
duration of the laser pulse there is the optimal density ($\omega
_{p}^{2}T^{2}=2$), at which the accelerating force peaks. When the laser
pulse propagates in inhomogeneous plasma the pulse duration will be
soon out of optimal value. In addition, the nonlinear dynamics of the
laser pulse during propagation can significantly
modify the pulse envelope and carrier frequency.
 The both effects influence the energy gain and should be
taken into account to find the plasma density profile for optimal
acceleration. 

The electron energy gain is $\Delta {\mathcal E} =\left( \int
F_{x}dx\right)  $. Introducing the parameter $\mu
=\int \left( n/n_{c}\right) dx$ the phase can be rewritten in the form
$\Phi =\mu \sqrt{{\dot{\mu} }}/2$ and the energy gain can be presented as
follows

\begin{eqnarray}
\Delta {\mathcal E}  &=&\int F(\xi ,\mu ,{\dot{\mu} })d\xi ,
\label{gain} \\
F(\xi ,\mu ,{\dot{\mu}}) &=& \frac{ \sqrt{\pi }} {2} a_{0}^{2}
{\dot{\mu} }\beta (\xi )\exp \left[ - \frac{\beta ^{2}(\xi ){\dot{\mu}
}}{4}\right] \cos \left( \frac{\mu \sqrt{{\dot{\mu} }}}{2} \right) , 
\label{force}
\end{eqnarray}

\noindent where the evolution of the pulse duration is assumed to be
known $\beta (\xi )=\omega T(\xi )$. Considering $\Delta {\mathcal E} \left[
\mu \left( \xi \right) \right] $ as a functional the Euler-Lagrange
equation can be derived for $\mu \left( \xi \right) $ providing the peak
gain  

\begin{equation}
\frac{\partial F}{\partial \mu }-\frac{d}{d\xi }\frac{\partial F}{\partial 
{\dot{\mu} }}=0.  
\label{Euler}
\end{equation}

\noindent Assuming that the laser pulse is short, $\omega _{p}T\ll 1$,
so that the pulse duration effects can be neglected $\beta (\xi )=0$,
Eq.~(\ref{Euler}) reduces to Eq.~(\ref{equation1}) with $\Phi
_{n}=-\pi (1+2n)$, where $n=1,2,...$  

Integrating Eq.~(\ref{equation1}) for homogeneous plasma
($n=n_{0}=\mbox{const}$) the wake phase at the bunch position can be
calculated $\Phi =-3\pi /2+(n_{0}/n_{c})^{3/2}\xi /2$, where it is
assumed that $\Phi (\xi =0)=-3\pi /2$. The electron acceleration is
possible in the range $0<x<L_{\hom }$, when $-\pi (3/2+2n)<\Phi <-\pi
(1/2+2n)$, where $n=1,2,...$ and $L_{\hom }=2\pi 
c(n_{0}/n_{c})^{-3/2}/\omega $ is the well known expression for
detuning length in homogeneous plasma \cite{Esarey1996}. Integrating
Eq.~(\ref{gain}) we obtain the known expression for the peak gain in the
electron energy for homogeneous plasma \cite{chen}
$\Delta {\mathcal E} _{\hom }=2\sqrt{2\pi }\exp
(-1/2)a_{0}^{2}(n_{c}/n_{0})mc^{2}$, where optimal duration of the
laser pulse $T=\sqrt{2}/\omega _{p}$ was assumed.  

For a short laser pulse $\omega _{p}T\ll 1$ the peak gain can be
achieved when the wake phase at the electron position is $\Phi
_{n}=-\pi (1+2n)$ and the plasma density profile obeys
Eq.~(\ref{solution1}). Integrating Eq.~(\ref{gain}) the energy gain
takes the form  

\begin{equation}
\Delta {\mathcal E} _{inh}(x)\simeq {\mathcal E} _{0}\left[ 1-\sqrt{\frac{n_{0}}{n(x)}}\right] ,
\label{ginh}
\end{equation}%

\noindent where ${\mathcal E} _{0}=(3/2)\sqrt{\pi } a_{0}^{2} mc^{2} (n_{0}/n_{c}) \left(
\omega T\right) (\omega L_{inh}/c)$. The energy gain over the distance
$0<x<L_{inh}$ is $\Delta {\mathcal 
E}_{inh}\simeq {\mathcal E}_{0}\simeq 
\Delta {\mathcal E} _{\hom } 2^{-3/2}\exp (1/2) \omega _{p0}T\Phi _{n}$, where
$\omega _{p0}^{2}=4\pi e^{2}n_{0}/m$. It follows from the obtained
expression that the gain increases as $\Phi _{n}$. Therefore, the
electron acceleration is most efficient when 
the electrons are loaded at the peak accelerating field as far behind
the laser pulse as possible. For arbitrary values of $\omega _{p}T$,
and the phase synchronism $\Phi_{n}= \mbox{const} $ ensured by the 
plasma profile (\ref{solution1}), the energy gain is

\begin{eqnarray} 
\Delta {\mathcal E} _{inh}\left( x\right)  &=& {\mathcal E} _{0}\left[
\Psi (x)-\Psi (0)\right] , 
\nonumber
\\
\Psi (x) &=&\sqrt{ \frac{\delta (0)}{\delta (x)}}\exp \left[ \delta (x) \right] + \sqrt{\pi
\delta (0)} \mbox{erf} \left[ \sqrt{\delta (x) } \right],
\label{ginh1}
\end{eqnarray}

\noindent
where $\delta (x)=\omega _{p}^{2}(x)T^{2}/4$ and $\delta (0)=\omega
_{p0}^{2}T^{2}/4=n_{0}\omega ^{2}T^{2}/4n_{c}$. In the limit $\omega
_{p}T\ll 1$ ($\delta \ll 1$) Eq.~(\ref{ginh1}) is reduced to
Eq.~(\ref{ginh}). 

\begin{figure}[tbp]
\includegraphics[width=6cm,clip]{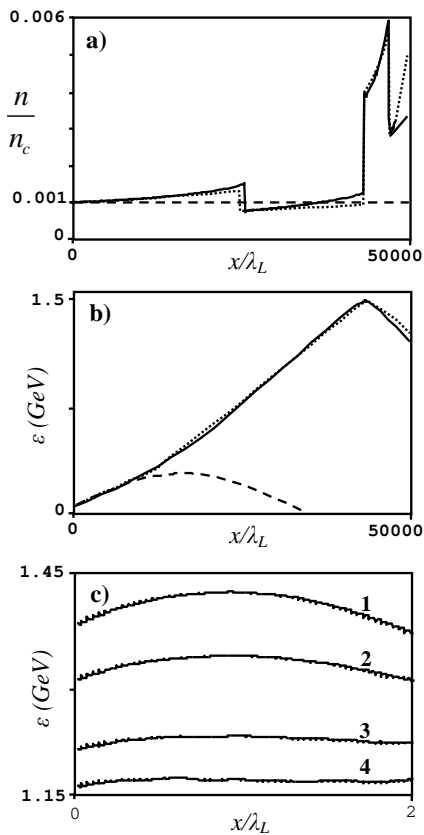}  
\caption{ Electron bunch acceleration and energy spread reduction in
plasma layers: (a) the plasma density profile, (b) the mean energy of
the electron bunch (c) electron energy vs electron position in
the bunch. Electrons are accelerated by the first laser pulse in
the first two layers ($0 < x < 43370 \lambda _L$) whereas they are
decelerated in the third and fourth layers
($43370 \lambda _L  < x < 49820 \lambda _L$). The solid line and the
dotted line correspond to the PIC simulation results and  the
theoretical estimates, respectively. To compare, the dashed line shows
acceleration in homogeneous plasma with the constant density 
$n_0 = 0.001 n_c$. The energy
distributions in frame (c) are shown at the beginning of the
deceleration at $x = 43370$~$\lambda _L$ (line 1), at $x =
45370$~$\lambda _L$  (line 2), at $x = 47370$~$\lambda _L$  (line 3)
and the end of deceleration at $x = 49820$~$\lambda _L$  (line 4),
respectively. The laser pulses are circularly polarized with Gaussian
profile and wavelength $1$~$\mu m$. The laser pulse parameters are
$T=10.6\lambda _{L}/c$, $a_{0}=0.6$ for the pulse on the
acceleration stage  and
$T=5.3\lambda _{L}/c$, $a_{0}=0.5$ for  the pulse on the
deceleration stage.  } 
\label{fig2}
\end{figure}

Keeping the bunch always in the same wave period behind the laser
pulse may become unpractical, because the plasma density would vary
too strongly over the full acceleration stage. To optimize the process
we propose a layered plasma density profile. The acceleration scheme is
illustrated in Fig.~\ref{fig1}. Let the electrons be trapped in the
$n$-th plasma period during acceleration in
the first plasma layer where the plasma density increases from $n_{1}$
to some final value $n_{f}$ in accordance with
Eq.~(\ref{solution1}). The wake phase is synchronized to the electron
bunch in this plasma layer due to the plasma density gradient. The
wake phase at the electron bunch center is $\Phi _{n}=\pi (2n-1)$ that
corresponds to the maximum of the accelerating force. When the plasma
density reaches the value of $n_{f}$, the electron bunch enters the
second plasma layer where the plasma density increases from $n_{2}$ to 
$n_{f}$. The density value of $n_{2}=n_{f}\Phi _{n-1}^{2}/\Phi
_{n}^{2}$ is chosen such that the electrons in the second layer
become located in the $(n-1)$-th plasma period behind the laser
pulse. The wake phase at the electron bunch center is $\Phi _{n-1}=\pi
(2n-3)$ that again corresponds to the maximum of the accelerating
force. Inductively, the plasma density increases from $n_{m}=n_{f}\Phi
_{n-m}^{2}/\Phi _{n-m-1}^{2}$ to $n_{f}$ in the $m$-th plasma layer in
accordance with Eq.~(\ref{solution1}) and the electrons are trapped
in the $\left( n-m \right) $-th plasma period behind the laser
pulse. The successive acceleration in the several plasma layers leads
to a significant increase in the electron energy gain and can be
accomplished by {\it the same driving laser pulse} until it is
depleted. The transition between the plasma density layers can be
smooth, but much shorter than the length of the layers themselves.

Further, the proposed scheme allows us to control the energy spectrum
of the bunch electrons. Naturally, electrons located at the bunch center get
the greatest energy gain while electrons located at the bunch
edges get the smallest energy gain. This leads to an energy
spread of the bunch after the acceleration stage. To reduce this
spread we suggest to use a matched decelerating stage with a higher
plasma density. Although the decelerating stage will reduce the net
energy gain, its effect on the energy spread reduction is much
stronger, because the plasma wake period decreases as the plasma
density increases.

To get an estimate of the energy spread reduction  we
assume that the electron bunch duration is much shorter than the
plasma wave period while the plasma density profile and energy gain
are described by Eqs.~(\ref{solution1}) and (\ref{ginh}),
respectively. It is also assumed that the bunch center is always
synchronized to the maximum of accelerating or decelerating forces
$\Phi _{n}=-\pi n$, $ n=1,2,...$ Expressing the phase $\Phi
_{n}$ in $\omega _{p}\eta/c\ll 1$, where $\eta$ is the distance from the
electron to the bunch center, the force acting at the electron located at the
position $\eta$ can be Taylor expanded as $F_{x}=F_{x0}(1-(\omega
_{p}\eta/c) ^{2}/2)$, where $F_{x0}$ is the force on the electron in the
bunch center. The difference in the accelerating force acting on
electrons located at different positions in the bunch leads to the
energy spread of the bunch electrons. Using Eq.~(\ref{ginh})
the energy gain as a function of the electron position in the bunch is 

\begin{eqnarray}
\Delta {\mathcal E} (x,\eta) &=& \Delta {\mathcal E} _{inh} \left( x \right) -\alpha \left( x \right) \eta^{2}
\\
\alpha \left( x \right) &=& {\mathcal E} _{0} \frac{\omega _{p0}^{2}}{2c^{2}}
\left[ 1-\sqrt{\frac{n}{n_{0}}} \right] ,
\label{gbunch}
\end{eqnarray}%
where $\Delta {\mathcal E} _{inh}$ is defined by Eq.~(\ref{ginh}).
\noindent Introducing the energy spread as $\sigma =\left(
\left\langle {\mathcal E} ^{2}\right\rangle -\left\langle {\mathcal E}
\right\rangle ^{2}\right) ^{1/2}$, where $\left\langle {\mathcal E}
\right\rangle =\int_{l_{b}/2}^{l_{b}/2}({\mathcal E}
/l_{b})d\eta$, $\left\langle {\mathcal E} ^2 \right\rangle
=\int_{l_{b}/2}^{l_{b}/2}({\mathcal E} ^{2}/l_{b})d\eta$ and $l_{b}/c$ is the 
bunch duration, we find the evolution of the bunch energy spread 

\begin{equation}
\sigma (x)=\frac{\alpha (x) l_{b}^{2}}{6\sqrt{5}}={\mathcal E}
_{0}\left[ \sqrt{\frac{n(x)}{n_{0}}} - 1 \right] \frac{\omega
_{p0}^{2}l_{b}^{2}}{12\sqrt{5}c^{2}}. 
\label{sigma}
\end{equation}

\noindent Let the bunch be accelerated in the first plasma layer where
the plasma density increases from $n_{1}$ to $4n_{1}$. Then, it is
decelerated in the second layer where the density increases from $n_{2}$ to
$4n_{2}$. The parameters of the laser pulse are assumed to be the same
in the both layers. The total energy gain after passing the two plasma
layers is ${\mathcal E} ={\mathcal E} _{1}[1-(\Phi _{2}/\Phi
_{1})(n_{1}/n_{2})^{1/2}]$ while the total spread is $\sigma =\sigma
_{1}|1-(\Phi _{2}/\Phi _{1})(n_{2}/n_{1})^{1/2}|$, where ${\mathcal E} _{1}
$ and $\sigma _{1}$ are the energy gain and spread after passing the
first layer,  
respectively,  $\Phi _{1}$ and $\Phi _{2}$ are the phase of the bunch
center in the first and second layers, respectively. Choosing
$n_{2}/n_{1}=\Phi _{1}^{2}/\Phi _{2}^{2}$ the final spread is removed
completely in this approximation. At the same time, the energy gain is 
${\mathcal E} ={\mathcal E} _{1}(1-n_{1}/n_{2})$. For $n_{2}/n_{1}
\simeq 4$ the bunch
loses only one quarter of its energy after passing the decelerating
layer while the energy spread will be completely removed. 

In order to check the validity of our simplified model, we have
performed 1D PIC simulations, using the code Virtual Laser Plasma
Laboratory \cite{vlpl}. The code has been supplemented with adaptive
scheme: the plasma density is varied in each time step so that the
bunch center is always located in the maximal value of the
accelerating force. The plasma wake excited by the first laser pulse
in the first two layers accelerates the electron bunches while the
bunch is decelerated in the second two layers where the plasma wake
excited by the second pulse (see Fig.~\ref{fig2}a). The laser pulses
are circularly polarized with Gaussian profile. The laser wavelength
$\lambda _{L}=1~\mu$m. 

The duration of the first pulse is
$T=10.6\lambda _{L}/c$ and $a_{0}=0.6$ while the duration of the
second pulse is $T=5.3\lambda _{L}/c$ and $a_{0}=0.5$. The plasma
density at the beginning of acceleration is $n_{0}=0.001\,n_{c}$. The
bunch center in the first layer is located at $3.5$ plasma periods
behind the pulse center (phase of bunch center is $\Phi _{1}=-7\pi$). 
The second layer is chosen such that the bunch is located at $2.5$
plasma periods behind the pulse center ($\Phi _{2}=-5\pi $) there. 
For the decelerating layers $\Phi _{3}=-8\pi $ and $\Phi _{4}=-6\pi $ in the
third and fourth layers. The electron bunch was initially
monoenergetic with energy $50$~MeV and duration
$T_{b}=2\lambda _{L}/c$.  

The bunch energy achieves about $1.4$~GeV with the energy spread of about
$2\%$ after acceleration in the first two plasma layers (see
Fig.~\ref{fig2}b,c). On the deceleration stage, the bunch energy
reduces to $1.2$~GeV whereas the energy spread reduces to less than
$0.5\%$ (see Fig.~\ref{fig2}b,c). The total distance
of the bunch acceleration and conditioning is about
$5$~cm. It is seen from Fig.~\ref{fig2}a,b that the plasma density
profile and the bunch energy evolution obtained in PIC simulation
agree fairly good with the theoretical predictions
(\ref{solution1}) and (\ref{ginh1}). To achieve a better agreement we 
take into account the pulse compression in the second half of the
first layer (we suppose that $a_{0}=0.7$ and $T=$ $7.8\lambda
_{L}/c$), in the second layer ($a_{0}=0.8$ and $T=$ $5.9 \lambda
_{L}/c$) and in the fourth layer ($a_{0}=0.6$ and $T=$ $3.7\lambda
_{L}/c$) in accordance with PIC simulation results. The deviation of
the numerical results from theoretical estimates is caused by the
complex nonlinear dynamics of the laser pulse during
propagation. According to Fig.~\ref{fig2}b, the energy gain in the
inhomogeneous plasma is about $5$ times more than that in the
homogeneous plasma with $n=n_0= \mbox{const}$.

The approach reported above is one-dimensional and thus does not take
into account transverse dynamics of the laser pulse and
electrons. The laser pulse can be efficiently guided by plasma
channels over many Rayleigh lengths \cite{Leemans1996,hooker}. More
complicated can be the transverse dynamics of accelerated
electrons. It is well known \cite{chen} that the simultaneous
accelerating and focusing of electrons occurs over a 
quarter of the plasma wavelength. Another accelerating quarter of the
plasma wave defocuses electrons. At the maximum of the accelerating
fields, however, the transverse fields vanish. The use of a preformed
plasma channel can significantly extend the region 
where the accelerating and focusing phases overlap \cite{Andreev1997}.

In conclusion, we have proposed to control both the maximum energy
gain and the accelerated beam quality by the plasma density
profile. Choosing a proper density gradient one can uplift the
dephasing limitation. The natural energy spread of
the particle bunch acquired at the acceleration stage can be
effectively removed by a matched deceleration stage, where a larger
plasma density is used.

\begin{acknowledgments}
This work has been supported in parts by Russian Foundation for Basic Research
(Grant No 07-02-01239) and by DFG Transregio TR-18, Germany. 
\end{acknowledgments}

\end{document}